\documentclass[pdftex,nofootinbib,superscriptaddress,aps,prfluids,notitlepage]{revtex4}
\usepackage{multirow,enumerate,epsfig,graphics,amssymb,amsmath,subeqnarray,mathrsfs,color,xcolor,epstopdf}
\usepackage{natbib,tikz}
\usetikzlibrary{calc}
\usepackage{graphicx}
\usepackage{dcolumn}
\usepackage{bm}
\usepackage{natbib}
\usepackage{color}
\usepackage{rotating}
\usepackage{amssymb}
\usepackage{amsmath}
\usepackage{gensymb}
\usepackage{booktabs}
\usepackage{bigints}
\usepackage{color,epsfig,graphics,amssymb,amsmath,subeqnarray,graphicx,amsthm,subfigure,mathrsfs}
\usepackage{mathtools,bm} 
\usepackage{rotating,mathrsfs}

\newcommand{\Reyn}{\mbox{Re}}
\newcommand{\sst}[1]{\scriptscriptstyle{#1}}
\newcommand{\oll}[1]{\overline{#1}}

\newcommand{\alp}{\alpha}
\newcommand{\veps}{\varepsilon}
\newcommand{\vecn}{\mbox{\bf n}}
\newcommand{\Id}{\mbox{\bf I}}
\newcommand{\veca}{\mbox{\boldmath$a$}}
\newcommand{\vecz}{\mbox{$\mathbf{e}_z$}}
\newcommand{\vecy}{\mbox{$\mathbf{e}_y$}}
\newcommand{\vecx}{\mbox{$\mathbf{e}_x$}}
\newcommand{\vecdel}{\mbox{\boldmath$\delta$}}
\newcommand{\veceps}{\mbox{\boldmath$\epsilon$}}
\newcommand{\vecOm}{\mbox{\boldmath$\Omega$}}
\newcommand{\dvecOm}{\;\;\dot{\mbox{\boldmath$\!\!\Omega$}}}
\newcommand{\vecom}{\mbox{\boldmath$\omega$}}
\newcommand{\vecu}{\mbox{\boldmath$u$}}
\newcommand{\vecv}{\mbox{\boldmath$v$}}
\newcommand{\vecU}{\mbox{\boldmath$U$}}
\newcommand{\vecUe}{\mbox{\boldmath$U$}_{\!e}}
\newcommand{\vecV}{\mbox{\boldmath$V$}}
\newcommand{\vecVO}{\mbox{\boldmath$V$}_{\!\!\sst O}}
\newcommand{\vecVG}{\mbox{\boldmath$V$}_{\!\!\sst G}}
\newcommand{\dvecVG}{\dot{\mbox{\boldmath$V$}}_{\!\!\sst G}}

\newcommand{\dvecV}{\dot{\mbox{\boldmath$V$}}}

\newcommand{\vecf}{\mbox{\boldmath$f$}}
\newcommand{\vecl}{\mbox{\boldmath$l$}}
\newcommand{\vecF}{\mbox{\boldmath$F$}}
\newcommand{\vecd}{\mbox{\boldmath$d$}}

\newcommand{\vecL}{\mbox{\boldmath$L$}}

\newcommand{\veck}{\mbox{\boldmath$k$}}
\newcommand{\vecr}{\mbox{\boldmath$r$}}

\newcommand{\veczero}{\mbox{\boldmath$0$}}
\newcommand{\vecsig}{\mbox{\boldmath$\sigma$}}

\newcommand{\Nab}{\mbox{\boldmath$\nabla$}}
\newcommand{\expo}{\mbox{e}}
\newcommand{\icomps}{\mbox{\scriptsize i}}
\newcommand{\icomp}{\mbox{i}}

\newcommand{\vecFp}{\mbox{\boldmath$F$}_{\! p}}

\newcommand{\vecLG}{\mbox{\boldmath$L$}_{\sst G}}

\newcommand{\vecfe}{\mbox{\boldmath$f$}_{\! e}}

\newcommand{\dsty}{\displaystyle}
\newcommand{\lam}{\lambda}
\newcommand{\tlam}{\tilde{\lambda}}
\newcommand{\fzero}{\;\hat{\!\!\vecf}_{\!\sst{0}}}

\makeatletter
\def\sgn{\mathop{\operator@font sgn}}
\def\threevdots{\vbox{\baselineskip1\p@ \lineskiplimit\z@
  \kern6\p@\hbox{.}\hbox{.}\hbox{.}}}
\makeatother

\begin{document}
\title{Acoustic propulsion of a small bottom-heavy sphere}
\author{Fran\c cois Nadal}
\email{F.R.Nadal@lboro.ac.uk}
\affiliation{Wolfson School of Mechanical, Electrical and Manufacturing Engineering,\\ Loughborough University,
LE11 3TU, Loughborough, United Kingdom}
\author{S\'ebastien Michelin}
\email{sebastien.michelin@ladhyx.polytechnique.fr}
\affiliation{LadHyX -- D\'epartement de M\'ecanique, CNRS -- Ecole Polytechnique, Institut Polytechnique de Paris, 91128 Palaiseau Cedex, France}

\begin{abstract}
We present here a comprehensive derivation for the speed of a small bottom-heavy sphere forced 
by a transverse acoustic field and thereby establish how density inhomogeneities may play a critical role in acoustic propulsion.
The sphere is trapped at the pressure node of a standing wave whose wavelength is much
larger than the sphere diameter. 
Due to its inhomogeneous density, the sphere oscillates in translation and rotation relative 
to the surrounding fluid. The perturbative flows
induced by the sphere's rotation and translation are shown to generate a rectified inertial 
flow responsible for a net mean force on the sphere that is able to propel the particle within the zero-pressure plane.
To avoid an explicit derivation of the streaming flow, the propulsion
speed is computed exactly using a suitable version of the Lorentz reciprocal theorem. 
The propulsion speed is shown to scale as the inverse of the viscosity, the cube of the amplitude of the acoustic field
and is a non trivial function of the acoustic frequency. Interestingly, for some combinations
of the constitutive parameters (fluid to solid density ratio, {\color{black} moment of inertia}
and centroid to {\color{black} center of mass} distance), the direction of propulsion is
reversed as soon as the frequency of the forcing acoustic field  becomes larger than a certain threshold.
The results produced by the model are compatible with both the observed phenomenology and the orders of magnitude
of the measured velocities.
\end{abstract}
\maketitle

\section{Introduction}

Controlled propulsion of microscopic objects in viscous flows has recently attracted much attention for
its potential biomedical applications such as drug transport and delivery \citep[]{Nelson2010, Sundararajan2008,Burdick2008}
or analytical sensing in biological media \citep[]{Campuzano2011b,Wu2010}.
Self-propulsion in viscous flows requires temporal and spatial symmetry-breaking~\citep{Purcell1977,Lauga2009}. 
Based on that principle, many different mechanisms have been proposed to achieve propulsion of small
rigid objects (see the reviews of Refs.~\citep{Ebbens2010,Wang2013}) and they generally belong to either of the two following categories.

The first and most classical group exploits an externally-applied directional field, that effectively breaks the
symmetry of the system at a scale much larger than the particle size, and drives the object in a specific direction. Electrophoresis 
\cite[]{Smoluchowsky1921} and diffusiophoresis \cite[]{Anderson1989} both fall in 
this first group, and result from the application of macroscopic electric or chemical gradients. The alternative approach relies on the local interaction of the particle with its close environment. Taking
advantage of its own asymmetry, the particle converts locally the energy provided by a  non-directional 
forcing field to break symmetry and self-propel.

For instance, catalytic
bimetallic micro-rods can propel themselves (self-electrophoresis) at high velocities (up to $10$ $\mu$m s$^{-1}$) by oxidizing 
hydrogen peroxide and exploiting the resulting self-generated local electric fields (see e.g.~\cite{Paxton2004,Ibele2007,Ebbens2011}).
For non-ionic solutes, the concentration gradient can also trigger
a net motion of the particle through self-diffusiophoresis \cite[]{Pavlick2011,Pavlick2013,Golestanian2007,Cordova2008}.
Similarly, autonomous propulsion can be achieved by taking advantage of self-thermophoresis effects
\cite[]{Jiang2010,Baraban2012,Qian2013}. Unfortunately, electrochemically- and thermally-based methods 
are not bio-compatible as a result of the inherent toxicity of the involved fuels (hydrogen peroxide, hydrazine) or of the required temperature differences. 

Alternatively, acoustic fields may be used to achieve autonomous motion in bio-fluids,
which explains the increasing interest of the scientific community in this type of propulsion method. \cite{Wang2012} demonstrated experimentally that bi-metallic
rods with asymmetric shape or composition were able to self-propel with velocities up to $200$ $\mu$m$\,$s$^{-1}$ when trapped in the nodal plane of an acoustic resonator. This pioneering work was soon extended to various configurations and geometries,  and self-acoustophoresis of 
magnetic clusters or asymmetric particles was thus reported~\citep{Sabrina2018,Ahmed2014,Ahmed2016}. Ref.~\cite{Kaynak2017} showed that
bio-inspired acoustic micro-swimmers with dedicated shapes were even able to reach
velocities up to 1200 $\mu$m$\,$s$^{-1}$. Although the prescribed acoustic field from which self-propulsion originates is directional, self-propulsion is achieved in a plane orthogonal to the excitation and its direction is not set by the external driving in constrast, for instance, with classical electrophoretic migrations of particles along the imposed forcing.

Since the seminal work of Ref.~\cite{Wang2012}, acoustic propulsion has been repeatedly ascribed to the streaming flows
self-generated by the particle's periodic motion with respect to its fluid environment of small yet finite inertia~\cite[]{Riley1966,Nadal2014, Collis2017, Kaynak2017}. 
To analyze the potential role of a particle's asymmetric shape on its ability to self-propel, Ref.~\cite{Nadal2014} first derived an integral form of the steady axial velocity of 
an acoustically-forced near-sphere, exploiting the absence of rotation of the particle at leading order in the particle's asymmetry as suggested by Ref.~\cite{Zhang1998}. Ref.~\cite{Lippera2018} recently showed however that this configuration did not actually yield any propulsion at leading order and that higher-order corrections in the particle's asymmetry were necessary to obtain a rectified effect. Ref.~\cite{Collis2017} considered the opposite case
of an asymmetric (in density or shape) dumbbell of large aspect ratio, and showed that the propelling streaming flow actually arose from the inertial coupling 
between the viscous flows respectively generated by the particle's translation and rotation, suggesting that acoustically-generated rotation of the particle was just as essential as its periodic translation in order to obtain acoustic propulsion.

Inspired by this observation, we analyse here how acoustic self-propulsion of a geometrically-symmetric particle (i.e. a sphere) may be achieved when its non-uniform density results in a combined translation and rotation under the effect of the acoustic forcing. We thus present the full analytical derivation of the leading order propulsion velocity of a non homogeneous sphere trapped
at the nodal plane of a resonator. The {\color{black} center of mass} and centroid of the
sphere do not coincide anymore, and as a result an inertial torque is imposed on the acoustically-forced sphere driving a combination of translational and rotational motions. We demonstrate that spherical particles may thus self-propel thanks to a symmetry-breaking in the hydrodynamic stress resulting from the inertial coupling of the viscous flows associated to the particle's combined translation and rotation~\cite{Collis2017}.

The paper is organized as follows. Section~\ref{sec:acoustic_forcing} is devoted to the derivation of the linear translational 
and rotational viscous responses of a non-homogeneous sphere to the transverse acoustic forcing.
The leading order inertial propulsion velocity of such a sphere is obtained in section~\ref{sec:sphere_prop} 
by means of a suitable version of the Lorentz reciprocal theorem. The physical relevance of this model to
experimental observations is then discussed in \S~\ref{sec:OM}. Finally, our main findings are summarised in \S~\ref{sec:conclusions}. 


\section{Acoustically forced dynamics of a sphere in a viscous fluid \label{sec:acoustic_forcing}}

\subsection{Configuration and main assumptions}
\begin{figure}
\begin{center}
\includegraphics[height=8cm]{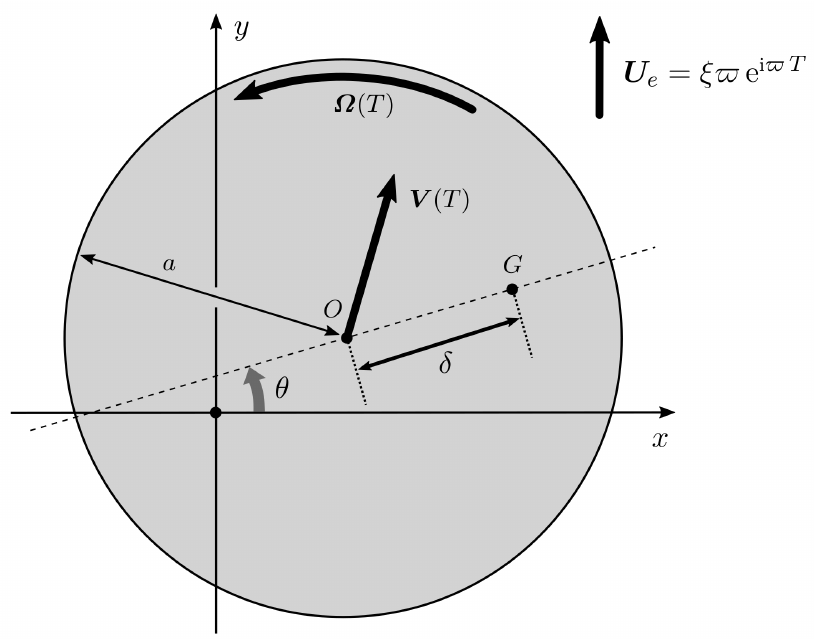}
\caption{Oscillations of a bottom-heavy sphere forced by a uniform external oscillating flow. {\color{black}$\xi$ and $\varpi$
are the displacement amplitude and frequency of the forcing acoustic field. The radius of the sphere is denoted by $a$,
and $\delta$ refers to the centroid-to-center of mass distance $OG$.}}\label{fig:geometry_config}
\end{center}
\end{figure}
We consider here the dynamics of a solid sphere of radius $a$ and mean density $\rho_s$ forced by a viscous periodic flow of density
$\rho$ and kinematic viscosity $\nu$. The density distribution of the solid sphere is not homogeneous, so that the 
center of mass $G$ departs from its centroid $O$ (Figure~\ref{fig:geometry_config}).
The mass and volume of the sphere are respectively 
$m_s = \rho_s\mathcal{V}_s$ and $\mathcal{V}_s = (4/3)\pi\,a^3$.

The forcing (acoustic) flow $\vecUe$ is uniform, harmonic of frequency $\varpi$ and directed 
along the $y$-direction. This configuration corresponds to a sphere of size $a$ trapped at the pressure node of a
standing acoustic wave of wave vector $\veck = k\,\vecy$ in the limit $ka \ll 1$
(with $(\vecx,\vecy,\vecz)$ are the Cartesian unit vectors). In such a case,
the incident flow can be considered as locally incompressible and to depend only on $y$.
We consider in the following that the external forcing flow is uniform and takes the
simple harmonic form
\begin{equation}
\vecUe = \hat{\vecU}_{\! e}\,\expo^{\icomps\varpi\,T} = \xi \varpi\,\,\expo^{\icomps\varpi\,T}\,\vecy,
\end{equation}
where $\xi$ is the amplitude of the fluid particles' displacement in the $y$-direction and is assumed to be much smaller 
than the particle's size $a$, so that $\veps=\xi/a\ll 1$. 

The offset of the sphere's centroid and center of mass is characterized by 
$\vecdel= \mathbf{OG}= \delta\,\vecd$, where $\vecd=\cos\theta\,\vecx+\sin\theta\,\vecy$ (Figure~\ref{fig:geometry_config}).
The velocity of $O$ and $G$ in the laboratory reference frame are denoted by 
$\vecVO$ and $\vecVG$ and the angular velocity $\vecOm$ of the sphere is aligned with $z$-direction: 
$\vecOm = \dot{\theta}\,\vecz$ (i.e. we assume that the sphere's density distribution is symmetric with respect to the $(Oxy)$-plane). In the following, we make the additional assumption that the velocity of the particle is periodic for the zero-mean forcing flow $\vecU_e(t)$ considered (note however, that its mean value is not necessarily zero so as to allow for self-propulsion regimes). 

The objective of the present section is to derive the response of the sphere to the external flow
in an unsteady Stokesian framework, where inertia of the fluid is negligible, but that of the solid particle is not.

\subsection{Momenta conservation}

The conservation of momentum in the (Galilean) frame of reference can be written
\begin{equation}
m_s\,\dvecVG = \vecF+\vecFp,
\label{eq:momentum1}
\end{equation}
where the total force experienced by the sphere is the sum of 
the hydrodynamic force $\vecF$ due to the relative velocity between the sphere and the surrounding fluid,
and of the pressure force $\vecFp = \rho \mathcal{V}_s\,\dot{\vecU}_e$ arising from the external pressure gradient that sets the fluid in motion. 
Such a distinction is justified by the form of the viscous drag experienced by a solid sphere oscillating in an uniformly oscillating flow
presented by Ref.~\cite{K&K}. Note that assuming that the velocity of the particle is periodic in time immediately implies that the time-average
of $\vecF(t)$ is zero.

Using $\vecVG = \vecVO - \vecdel \times \vecOm$, and noting $\vecV = \vecVO - \vecUe$ the velocity of the sphere relative to the
oscillating fluid, the previous equation can be rewritten
%
%
%
\begin{equation}
\rho_s\mathcal{V}_s\dvecV = m_s\left[(\vecOm \times \vecdel)\times\vecOm + \,\vecdel\times\dvecOm\right] +
\vecF + (\rho-\rho_s)\mathcal{V}_s\dot{\vecU}_e,
\label{eq:momentum3}
\end{equation}
where the last term is the effective buoyancy force.

Similarly, the conservation of  angular momentum can be written about the center of mass $G$, 
\begin{equation}
I_{\sst G}\,\dvecOm = \vecLG,
\label{eq:ang_momentum1} 
\end{equation}
where $I_{\sst G}$ is the {\color{black} moment of inertia} of the sphere about the $(G,z)$-axis and
$\vecLG$ is the total torque experienced by the sphere at its {\color{black} center of mass}. 

The sphere is rigid, thus $\vecLG = \vecL +  \mbox{\boldmath$GO$} \times (\vecF+\vecF_p)$, with $\vecL$ is the hydrodynamic torque 
about the geometric center $O$, and Eq.~\eqref{eq:ang_momentum1} finally becomes
%
%
%
\begin{equation}
I_{\sst G}\,\dvecOm =  \vecL -  \vecdel \times (\vecF + \vecFp).
\label{eq:ang_momentum3} 
\end{equation}
\subsection{Dimensionless forms of the conservation laws}
In the following, using $U_e = \xi \varpi$ 
and $\varpi^{-1}$ as reference velocity and time scales, respectively, 
yields the non-dimensional form of Eqs.~\eqref{eq:momentum3} and \eqref{eq:ang_momentum3} (using lower-case letters for dimensionless variables)
\begin{gather}
\dot{\vecv} = 
\veps^{-1}\,\alpha\,\left[(\vecom \times \vecd)\times\vecom + \vecd\times\dot{\vecom}\right] 
+ \left(\frac{3}{4\pi}\frac{\beta}{\lambda^2}\right)\,\vecf + (\beta -1)\,\vecfe, \label{eq:momentum4}\\
\dot{\vecom} = \left(\frac{15}{8\pi}\frac{\beta}{\lambda^2 I}\right)\,\vecl 
- \veps\left[\left(\frac{15}{8\pi}\frac{\alp\beta}{\lambda^2 I}\right)\,(\vecd \times \vecf)
+ \left(\frac{5}{2}\frac{\alp\beta }{I}\right)\,(\vecd \times \vecfe)\right], \label{eq:ang_momentum2}
\end{gather}
with $\vecfe = \icomp\,\expo^{\icomps t}\,\vecy$ the fluctuating forcing.\\

In the above equations, the dimensionless {\color{black} moment of inertia} $I = I_{\sst G}/I_{\sst 0}$ is the ratio between the
actual {\color{black} moment of inertia} $I_{\sst G}$ and $I_{\sst 0} = (2/5)\,m_s\,a^2$, the {\color{black} moment of inertia} of a homogeneous 
sphere with the same mean density with respect to its centre. Further, $\alpha=\delta/a$ is the relative geometric offset of the 
particle's {\color{black}center of mass} and thus characterizes its non-homogeneity, 
$\beta = \rho/\rho_s$ is the fluid-to-solid average density ratio and $\lam = (a^2\varpi/\nu)^{1/2}$ is
the ratio between the radius of the sphere and the viscous penetration length (i.e. $\lambda^2$ is the reduced frequency of actuation).
It should be noted that the $\veps^{-1}$ factors are associated with the choice of characteristic velocity scale. 
\subsection{Harmonic response in the {\color{black} unsteady Stokes limit} \label{subsec:lin_resp}}

The non-dimensional forcing field $\vecfe = f_e\,\vecy$ with $f_e = \icomp \expo^{\icomps t}$ is $O(\veps^0)$ and harmonic;
as a result, for $\veps\ll 1$, the leading order dynamics is obtained by noting that $v_y = O(1)$ while $v_x$ and $\theta$ are $O(\veps)$: 
%
\begin{gather}
\dot{v}_x =  \alpha \veps^{-1}\,(\dot{\theta}^2 + \ddot{\theta}\theta) 
+ \left(\frac{3}{4\pi}\frac{\beta}{\lambda^2}\right)\,f_{x},\label{eq:momentumx}\\
\dot{v}_y = - \alpha\veps^{-1}\,\ddot{\theta} 
+ \left(\frac{3}{4\pi}\frac{\beta}{\lambda^2}\right)\,f_{y} + (\beta-1)\,f_e \label{eq:momentumy},\\
\ddot{\theta} = \left(\frac{15}{8\pi}\frac{\beta}{\lambda^2 I}\right)\,l 
- \veps\left[\left(\frac{15}{8\pi}\frac{\alp\beta}{\lambda^2 I}\right)\,f_{y} 
+\left(\frac{5}{2}\frac{\alp\beta }{I}\right)\,f_e\right]. \label{eq:ang_momentumz}
\end{gather}
%
In the {\color{black} unsteady Stokes limit}, the total viscous  force and 
torque on the sphere are obtained by superimposing that induced by the sphere's translation and
rotation independently. By symmetry, the force induced by the sphere's rotation and the torque 
(about $O$) induced by the sphere's translation are both identically zero. Therefore, considering the
form of the system \eqref{eq:momentumx}--\eqref{eq:ang_momentumz} and according to Ref.~\cite{K&K},
one can write
\begin{gather}
\dsty v_x = \veps\,\hat{v}_{{\sst 0},x}\,\expo^{2\icomps  t},\;\;\dsty v_y = \hat{v}_{{\sst 0},y}\,\expo^{\icomps t},
\;\;\theta = \veps\,\hat{\theta}_{\sst 0} \,\expo^{\icomps t}
\label{eq:velocities_eps_freq}
\end{gather}
and
\begin{gather}
f_{x} = - \Delta_{\sst 2}\,v_x,\;\;f_{y} = - \Delta_{\sst 1}\,v_y,\;\;l = -\Lambda_{\sst 1}\,\dot{\theta},
\label{eq:torque_force_eps_freq}
\end{gather}
where the drag coefficients $\Delta_n$ and $\Lambda_n$ are associated with harmonic translational or rotational motion of a sphere in unsteady viscous flows \citep{K&K} (see also \S~\ref{subsec:steady_prop}).
\begin{equation}
\Delta_n = 6\pi\left(1 + \,n^{1/2}\tlam + \frac{n \tlam^2}{9}\right) \;\;\mbox{and} \;\;
\Lambda_n = 8\pi\,\frac{1 + n^{1/2}\tlam + \icomp\, n \tlam^2/3}{1 + n^{1/2}\tlam},
\label{eq:Del_Lam}
\end{equation}
and $\tlam = \expo^{\icomps \pi/4}\,\lambda$. 

The leading order dynamics ($\hat{v}_{{\sst 0},y}$,$\hat{\theta}_{\sst 0}$) is then obtained from the linear system:
\begin{gather}
\left[\icomp + \left(\frac{3}{4\pi}\frac{\beta}{\lambda^2}\right)\,\Delta_{1}\right]\hat{v}_{{\sst 0},y} 
- \alpha\,\hat{\theta}_{\sst 0}
= \icomp (\beta - 1),  \label{eq:lin_sys_vy}\\
\left(\frac{15}{8\pi}\frac{\alp\beta}{\lambda^2 I}\right)\,\Delta_{1} \hat{v}_{{\sst 0},y} 
+ \left[1 - \icomp\left(\frac{15}{8\pi}\frac{\beta}{\lambda^2 I}\right)\,\Lambda_{1}\right]\,\hat{\theta}_{\sst 0}
= \icomp \left(\frac{5}{2}\frac{\alp\beta }{I}\right). 
\label{eq:lin_sys_theta}
\end{gather}
It should be noted that the above dynamics is independent from that along the $x$-direction, which can be computed in a second step.
The complex amplitude of the angular velocity $\hat{\omega}_{\sst 0}$ is then obtained using 
$\hat{\omega}_{\sst 0} = \icomp\,\veps\,\hat{\theta}_{\sst 0}$. 


From the small and large $\lambda$ approximations of $\Delta_1$ and $\Lambda_1$, Eqs.~\eqref{eq:lin_sys_vy}--\eqref{eq:lin_sys_theta} 
can be used to obtain the following useful asymptotic forms of $\hat{v}_{{\sst 0},y}$ and $\hat{\theta}_{\sst 0}$:
\begin{align}
\hat{v}_{{\sst 0},y}\sim\frac{2\icomp\lambda^2(\beta-1)}{9\beta}\;\;&\mbox{and}\;\; \hat{\theta}_{\sst 0}\sim-\frac{\alpha\lambda^2}{6\beta}
\;\;\mbox{for}\;\;\lambda\rightarrow 0,\label{eq:asymp_small_lambda}\\
\hat{v}_{{\sst 0},y}\sim\frac{4I(\beta-1)+10\alpha^2\beta}{\displaystyle 4I+\beta(2I+5\alpha^2)}\;\;&\mbox{and}\;\;\hat{\theta}_{\sst 0}\sim\frac{\displaystyle\displaystyle 15\icomp\alpha\beta}{4I+\beta(2I+5\alpha^2)}
\;\;\mbox{for}\;\;\lambda\rightarrow\infty.\label{eq:asymp_large_lambda}
\end{align}
%

\section{Acoustic propulsion of the sphere \label{sec:sphere_prop}}

Knowing the leading-order viscous response of the sphere to the incident acoustic field, we now proceed to explore the possibility to
achieve propulsion by means of streaming effects, by accounting for the first inertial correction to the flow field following the approach 
of Ref.~\cite{Lippera2018}.


\subsection{Governing equations \label{subsec:geom_gov}}

By moving through the fluid, the sphere generates a flow field $\vecu(\vecr,t)$ around itself governed by the
Navier-Stokes and continuity equations, which can be written in non-dimensional form in the frame of reference moving with the fluid far from the sphere as
\begin{equation}
\dsty \lam^2\frac{\partial \vecu}{\partial t}+ \Reyn\,\Nab\vecu\cdot\vecu = \Nab\cdot\vecsig,\;\; 
\Nab\cdot\vecu = 0,
\label{eq:NS_adim}
\end{equation}
%
%
where $\vecsig$ is the non-dimensional hydrodynamic stress in the fluid due to the relative
motion between the sphere and the surrounding fluid (and therefore includes a corrected pressure to account for the inertial corrections associated with the moving frame). It is recalled that, as in the previous section, all quantities are non-dimensional and $a$, $\varpi^{-1}$, and $\xi\varpi$ are used as reference length, time and velocity scales respectively. In Eq.~\eqref{eq:NS_adim}, the Reynolds number is $\Reyn=\veps\lam^2$
with $\veps = \xi/a\ll 1$. In the following we thus focus on the limit of $\Reyn \ll 1$, which yields the restriction $\lambda^2\ll \veps^{-1}$ for the following analysis. {\color{black} Note that $\Reyn$ is therefore not a new independent dimensionless group so that the problem is only governed by 
the five parameters $\veps$, $\lam$, $\alpha$, $\beta$ and $I$ defined in the previous section, and listed in table \ref{tab:dimless_param}.}
\begin{table}
\begin{center}
\color{black}
\begin{tabular}{ccc}
{\bf parameters} 	& {\bf expression} 			& {\bf physical meaning} \\ \midrule
$\beta$				& $\rho/\rho_s$				& fluid-to-solid density ratio\\
$\varepsilon$ 		& $\xi/a$             		& dimensionless displacement amplitude of the acoustic field \\
$\alpha$      		& $\delta/a$          		& dimensionless imbalance parameter \\
$I$           		& $I_{\sst G}/I_{\sst 0}$ 	& dimensionless moment of inertia \\
$\lambda$ 			& $(\varpi a^2/\nu)^{1/2}$ 	& inverse of the dimensionless viscous length\\
\end{tabular}
\caption{List of the five independent parameters of the problem. Note that the Reynolds number $\Reyn = \varepsilon \lambda^2$, which is
supposed to be small compared to unity, is not an independent parameter.\label{tab:dimless_param}}
\end{center}
\end{table}

The flow field vanishes at infinity and satisfies the no-slip boundary condition on the moving sphere  ($\vert \vecr \vert = 1$ in a set of axes attached to the centroid of the sphere), therefore
\begin{equation}
\vecu = \vecv + \vecom \times \vecr \;\; \mbox{for}\; \vert \vecr \vert = 1,\qquad 
\vecu \rightarrow 0 \;\;\mbox{for}\; \vert \vecr \vert\rightarrow \infty.\label{eq:BC12}
\end{equation}

\subsection{Expansions in power of $\Reyn$ and order of the propulsion speed}

The Reynolds number $\Reyn$ is a small 
parameter of the problem, and we now expand the velocity field $\vecu$, the hydrodynamic stress
$\vecsig$, and the {\color{black} velocity} of the sphere $\vecv$ in powers of the Reynolds number
\begin{equation}
\vecu = \vecu^{\sst{(0)}} + \Reyn\,\vecu^{\sst{(1)}} + \cdots,\qquad 
\vecsig = \vecsig^{\sst{(0)}} + \Reyn\,\vecsig^{\sst{(1)}} + \cdots,\qquad 
\vecv = \vecv^{\sst{(0)}} + \Reyn\,\vecv^{\sst{(1)}} + \cdots. \label{eq:expans_v}
\end{equation}
We are interested in the emergence of a net propulsion of the sphere and therefore will focus on the existence of a steady component to the sphere's velocity. 
Due to the linearity of the unsteady Stokes equation, such steady motions have to be 
generated at order $O(\Reyn)$ at least, which can be written $\oll{\vecv} = \Reyn\,\langle\vecv^{\sst{(1)}}\rangle$,
where $\langle\cdots\rangle$ refers to the time average operator over a period of oscillation.
In other words, the possibly non-zero $O(\Reyn)$ steady  component of the 
speed $\oll{\vecv}$ must be induced by the steady 
streaming flow resulting from the self-coupling of the $O(1)$ (i.e. $\Reyn = 0$) viscous flow through the 
nonlinear term of the Navier-Stokes equation. 

To obtain such a forcing, one could explicitly derive the steady streaming flow and integrate the 
corresponding hydrodynamic stress over the surface of the sphere.
In order to circumvent such a cumbersome derivation, we use in the following 
a specific form of Lorentz reciprocal theorem suitable for the case where inertial 
corrections are considered \citep{ho_leal_1974,Nadal2014,Lippera2018}.  

\subsection{Lorentz reciprocal theorem for inertial corrections}

\color{black} To this end, we define the auxiliary flow and stress fields $(\vecu^\star,\vecsig^\star)$, as the unique solution of the following \emph{steady} Stokes problem
\begin{equation}
{\color{black} \Nab\cdot\vecsig^\star = \veczero} \;\;\mbox{and}\;\;\Nab\cdot\vecu^\star = 0,\label{eq:Stokes_star}
\end{equation}
with boundary conditions
\begin{equation}
\vecu^\star = \vecv^\star + \vecom^\star \times \vecr \;\; \mbox{at}\;\vert\vecr\vert = 1,\qquad
\vecu^\star \rightarrow 0 \;\;\mbox{for}\; \vert \vecr \vert\rightarrow \infty.
\label{eq:BC_star}
\end{equation}

Using Eqs.~\eqref{eq:NS_adim} and \eqref{eq:Stokes_star} and denoting by $\mathcal{V}$ the volume of fluid outside the sphere,
one can write an instantaneous version of the Lorentz reciprocal theorem (for further details, see again Ref.~\citep[][]{Lippera2018}) in the following form:
%
\begin{equation}
\lam^2\int_{\sst{\mathcal{V}}}\vecu^\star\cdot\frac{\partial \vecu}{\partial t}\,d\mathcal{V}
+ \Reyn \int_{\sst{\mathcal{V}}}[\vecu^\star\cdot\Nab\vecu\cdot\vecu]\,d\mathcal{V} = 
\vecf^\star \cdot \vecv + \vecl^\star \cdot \vecom - \vecv^\star \cdot \vecf - \vecom^\star \cdot \vecl.
\label{eq:Lorentz3}
\end{equation}
where $\vecf$ and $\vecl$ (resp. $\vecf^\star$ and $\vecl^\star$) are the hydrodynamic force and 
torque in $O$ for the real (resp. auxiliary) problem. 
Because the particle is spherical, we immediately have $\vecf^\star = -6\pi\,\vecv^\star$ and $\vecl^\star = -8\pi\,\vecom^\star$
and Eq.~\eqref{eq:Lorentz3} becomes
\begin{equation}
-(6\pi\vecv+\vecf)\cdot\,\vecv^\star  -(8\pi \vecom+\vecl)\cdot \vecom^\star  = \lam^2\frac{d}{dt}\left[\int_{\sst{\mathcal{V}}}\vecu^\star\cdot \vecu\,d\mathcal{V}\right]
+ \Reyn \int_{\sst{\mathcal{V}}}[\vecu^\star\cdot\Nab\vecu\cdot\vecu]\,d\mathcal{V},
\label{eq:lorenz4}
\end{equation}
since $\vecu^\star$ is time-independent 
and $\mathcal{V}$ is fixed in time. %
It should be noted that up until now, no assumption on the magnitude of $\Reyn$ was used and the 
previous equation is therefore valid for any 
value of the Reynolds number.\\

Now, introducing Eqs.~\eqref{eq:expans_v} and the additional $\Reyn$-expansions
\begin{equation}
\vecf = \vecf^{\sst{(0)}} + \Reyn \vecf^{\sst{(1)}} + \cdots,\qquad 
\vecl = \vecl^{\sst{(0)}} + \Reyn\,\vecl^{\sst{(1)}} + \cdots.\label{eq:expans_l}
\end{equation}
%
%
for the force and torque into Eq.~\eqref{eq:lorenz4}, and
and identifying  the $O(1)$  terms, leads to
\begin{equation}
-(6\pi\vecv^{\sst{(0)}}+\vecf^{\sst{(0)}})\cdot\,\vecv^\star  -(8\pi \vecom^{\sst{(0)}}+\vecl^{\sst{(0)}})\cdot 
\vecom^\star  =
\lam^2\int_{\sst{\mathcal{V}}}\vecu^\star\cdot\frac{\partial \vecu^{\sst{(0)}}}{\partial t}\,d\mathcal{V}.
\label{eq:lorenz_O1}
\end{equation}
%
Note that the right-hand side of Eq.~\eqref{eq:lorenz_O1} can be integrated provided assumptions
on the harmonic nature of the $O(1)$ solution are formulated, in order to obtain the drag force and torque 
in unsteady Stokes flow ($\Reyn=0$,
see \S~\ref{subsec:steady_prop}).    \\


%
Considering now the $O(\Reyn)$ terms in Eq.~\eqref{eq:lorenz4}, the problem obtained at that order is structurally similar to that at $O(1)$ but for the emergence of an extra forcing that arises from and accounts for the effect of the streaming flow. 
%
%
Should a net self-propulsion occur (i.e. on average 
over a whole period of forcing), it would therefore be due to the streaming forcing, as anticipated. 
Taking the average in time of the resulting equation, one obtains
\begin{equation}
-\vecv^\star \cdot \langle 6\pi\,\vecv^{\sst{(1)}}+\vecf^{\sst{(1)}}\rangle 
-\,\vecom^\star \cdot \langle 8\pi\vecom^{\sst(1)}+\vecl^{\sst{(1)}} \rangle  =
\left\langle \int_{\sst{\mathcal{V}}}[\vecu^\star 
\cdot\Nab\vecu^{\sst{(0)}}\cdot\vecu^{\sst{(0)}}]\,d\mathcal{V}\right\rangle=\mathcal{H}.
\label{eq:lorenz2_ORe}
\end{equation}
\color{black}

In order to derive the steady component of the propulsion speed $\oll{\vecv} = 
\Reyn\,\langle \vecv^{\sst{(1)}}\rangle$, our goal in the following lies in the computation of the right-hand-side, 
$\mathcal{H}$, of the previous equality. 

\subsection{Viscous drags and steady propulsion speed  \label{subsec:steady_prop}}

Knowing the form of the viscous dynamical response of the sphere
$(\hat{\vecv}_{\sst 0},\hat{\omega}_{\sst 0})$ 
from \S~\ref{sec:acoustic_forcing}, we are now able to derive 
an explicit expression of the propulsion speed $\bar{\vecv}$. We first write 
$\vecv^{\sst{(0)}} = \hat{\vecv}_{\sst{0}}\expo^{\icomps t}$, 
$\vecom^{\sst{(0)}} = \hat{\vecom}_{\sst{0}}\expo^{\icomps t}$, 
$\vecu^{\sst{(0)}} = \hat{\vecu}_{\sst{0}}\expo^{\icomps t}$, 
$\vecf^{\sst{(0)}} = \fzero\expo^{\icomps t}$ and 
$\vecl^{\sst{(0)}} = \hat{\vecl}_{\sst{0}}\expo^{\icomps t}$.

In this context, the $O(1)$ and $O(\Reyn)$ components of Eqs.~\eqref{eq:lorenz_O1} and \eqref{eq:lorenz2_ORe} become
\begin{align}
- 6\pi(\hat{\vecv}_{\sst{0}}+\fzero)\cdot\vecv^\star -(8\pi\hat{\vecom}_{\sst0}+\hat{\vecl}_{\sst{0}})\cdot\vecom^\star
  &=
\lam^2\int_{\sst{\mathcal{V}}} \vecu^\star\cdot\frac{\partial \hat{\vecu}_{\sst{0}}}{\partial t}
\,d\mathcal{V}
\label{eq:lorenz_O1_harmonic}\\
-\vecv^\star \cdot \langle 6\pi\,\vecv^{\sst{(1)}}+\vecf^{\sst{(1)}}\rangle 
-\,\vecom^\star \cdot \langle 8\pi\vecom^{\sst(1)}+\vecl^{\sst{(1)}} \rangle &=
\frac{1}{2}
\Re\left\{\int_{\sst{\mathcal{V}}}
[\vecu^\star \cdot\Nab\hat{\vecu}_{\sst{0}}^\dag\cdot\hat{\vecu}_{\sst{0}}]\,d\mathcal{V}\right\}
=\mathcal{H}.
\label{eq:lorenz_ORe_harmonic}
\end{align}
where $\Re(z)$ and $z^\dag$ stand for the real part and complex conjugate of $z$.

In the case of an harmonic motion, $\hat{\vecu}_{\sst{0}}$ is given by
\begin{equation}
\hat{\vecu}_{\sst{0}} =  \left[A(r)\Id + B(r)\vecn\vecn\right]\cdot\hat{\vecv}_{\sst 0}
+ C(r)\,\hat{\vecom}_{\sst{0}}\times\vecn,
\label{eq:harmonic_viscous_field}
\end{equation}
where the exact forms for given $\lambda$ of $A(r)$, $B(r)$ and $C(r)$ are reminded in Appendix~\ref{app:viscous_steady_harmonic} (see also
chapter 6 in Ref.~\citep[][]{K&K}). The velocity field induced by a rectilinear
steady motion of a sphere in a viscous fluid has a form similar to Eq.~\eqref{eq:harmonic_viscous_field}
\begin{equation}
\vecu^\star =  \left[A^\star(r)\Id + B^\star(r)\vecn\vecn\right]\cdot\vecv^\star
+ C^\star(r)\,\vecom^\star\times\vecn,
\label{eq:steady_viscous_field}
\end{equation}
where the exact forms of $A^\star(r)$, $B^\star(r)$ and $C^\star(r)$ are also given in appendix 
\ref{app:viscous_steady_harmonic}, and are in fact respectively the asymptotic limits of $A$, $B$
and $C$ for $\lambda\rightarrow 0$ (steady motion).

\subsubsection{Order $O(1)$ - Viscous response}

Successively introducing the auxiliary fields 
$(\vecv^\star,\vecom^\star) = (\vecy,\veczero)$ and $(\vecv^\star,\vecom^\star) = (\veczero,\vecz)$
in Eq.~\eqref{eq:lorenz_O1_harmonic} provides

\begin{align}
\fzero   &= -[6\pi+\tlam^2 F(\lambda)]\,\hat{\vecv}_{\sst 0},\qquad 
\hat{\vecl}_{\sst 0}  = -[8\pi+\tlam^2 G(\lambda)]\,\hat{\vecom}_{\sst 0},\label{eq:order0torque}
\end{align}
with 
\begin{equation}
F(\lambda) = 4\pi\int_1^\infty r^2\left[\frac{2AA^\star+(A+B)(A^\star+B^\star)}{3}\right]dr,\quad
G(\lambda) = \frac{8\pi}{3}\int_1^\infty r^2 CC^\star\,dr,
\label{eq:FG_lambda}
\end{equation}
and $\tlam^2 = \icomp\lambda^2$.
One can note that computing the integral on the right-hand sides of Eqs.~\eqref{eq:FG_lambda}
indeed provides the classical expressions
derived for the unsteady translational and rotational drag \citep{Stokes1850,Mazur1974,K&K}: 
\begin{equation}
\fzero  = -6\pi\left(1 + \tlam + \frac{\tlam^2}{9}\right)\hat{\vecv}_{\sst 0},\qquad
\hat{\vecl}_{\sst 0}  = -8\pi \frac{1+\tlam+\tlam^2/3}{1+\tlam}\,\hat{\vecom}_{\sst 0}.\label{eq:order0torque_int}
\end{equation}

\subsubsection{Order $O(\Reyn)$ - Propulsion speed}

\begin{figure}
\begin{center}
\includegraphics[width=.95\textwidth]{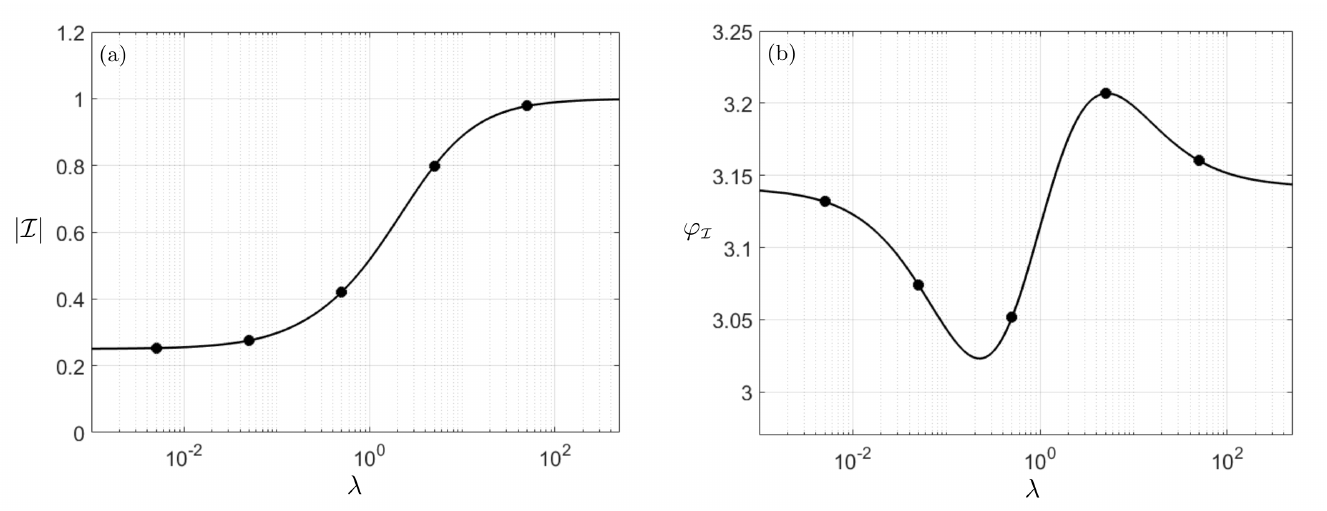}
\caption{Magnitude (a) and phase (b) of $\mathcal{I}$. The bullets correspond to the direct
numerical integration of Eq.~\eqref{eq:I}.}
\label{fig:I}
\end{center}
\end{figure}

Let us turn to the leading-order mean propulsion speed $\oll{\vecv}$. 
Using Eq.~\eqref{eq:harmonic_viscous_field}, 
\begin{align}
\Nab\hat{\vecu}_{\sst 0} = A'\hat{\vecv}_{\sst 0}\vecn 
+ B'(\vecn\cdot\hat{\vecv}_{\sst 0})\vecn\vecn\,
+ &\,\frac{B}{r}\left\{(\vecn\cdot\hat{\vecv}_{\sst 0})(\mathbf{I}-\vecn\vecn)
+ \vecn\otimes[(\mathbf{I}-\vecn\vecn)\cdot\hat{\vecv}_{\sst 0}]\right\}\notag\\
&+  C'(\hat{\vecom}_{\sst 0}\times\vecn)\otimes\vecn
-\frac{C}{r}\left[\veceps\cdot\hat{\vecom}_{\sst 0}+(\hat{\vecom}_{\sst 0}
\times\vecn)\otimes\vecn\right],
\label{eq:grad_u0}
\end{align}
where $C' = dC/dr$ and $(\veceps)_{ijk} = \epsilon_{ijk}$, so that $(\veceps\cdot\vecom_{\sst 0})\cdot \veca 
= \veca \times \vecom_{\sst 0}$ for any vector $\veca$. 
Introducing Eqs.~\eqref{eq:harmonic_viscous_field}, \eqref{eq:steady_viscous_field} 
and \eqref{eq:grad_u0} in Eq.~\eqref{eq:lorenz_ORe_harmonic} and performing the explicit integration
of its right-hand side leads to
\begin{equation}
\mathcal{H} = \frac{2\pi}{3}\Re
\bigg\{\left[\vecv^\star\cdot(\hat{\vecom}_{\sst{0}}\times\hat{\vecv}_{\sst{0}}^\dag)\right]
\mathcal{I}(\tlam)\bigg\},
\label{eq:H}
\end{equation}
where the quantity 
\begin{equation}
\mathcal{I}(\tlam) = 
\bigintsss_1^\infty\bigg[A^\star\bigg(A^\dag C'+B^\dag C'+ \frac{2A^\dag C}{r}\bigg)
+ \frac{B^\star(A^\dag-B^\dag) C}{r}\bigg]r^2\,dr.
\label{eq:I}
\end{equation}
is given in its fully integrated form in appendix \ref{app:I} and its variations are indicated on Figure~\ref{fig:I}. 
In particular, for small and large $\lambda$, the asymptotic behaviour of $\mathcal{I}$ is obtained as
\begin{equation}
\mathcal{I}(\lambda\rightarrow 0)=-\frac{1}{4},\qquad \mathcal{I}(\lambda\rightarrow \infty)=-1.\label{eq:asymp_I}
\end{equation}
\begin{figure}
\begin{center}
\includegraphics[width=.95\textwidth]{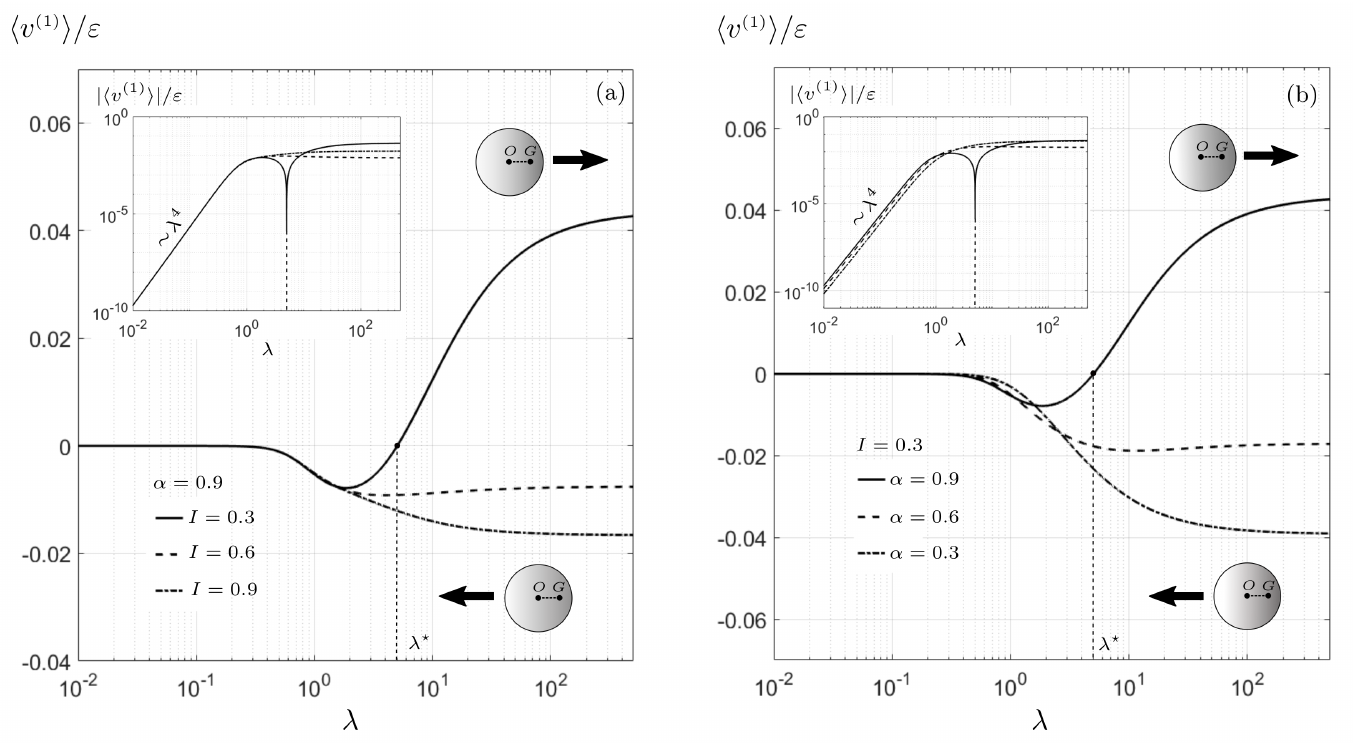}
\caption{Ratio $\langle v^{\sst(1)} \rangle/\veps$ as a function of $\lambda$ for $\beta = 0.2$ and
several combinations of $(I,\alpha)$. (a) $\alpha = 0.9$, $I = 0.3,\,0.6,\,0.9$ ; (b) $I = 0.3$, 
$\alpha = 0.3,\,0.6,\,0.9$. The spheres sketched in (a) and (b) illustrate the direction of propulsion
when the {\color{black} center of mass} is on the right of the geometric center (top: $\langle v^{\sst(1)} \rangle/\veps > 0$ ; 
bottom: $\langle v^{\sst(1)} \rangle/\veps < 0$).}
\label{fig:H_vs_lambda}
\end{center}
\end{figure}
Note that Eq.~\eqref{eq:H} confirms that the $x$-component of $\hat{\vecv}_{\sst 0}$ will have no
contribution to the steady motion, as anticipated in \S~\ref{sec:acoustic_forcing} and expected for symmetry reasons.

Now, choosing $\vecv^\star=\vecx$ and $\vecom^\star=\veczero$ in Eqs.~\eqref{eq:lorenz_ORe_harmonic} and \eqref{eq:H}, and remembering that (i) using Eq.~\eqref{eq:velocities_eps_freq}, only the $y$-component of $\hat{\vecv}_{\sst 0}$ has a non zero contribution to the
mean propulsion speed, (ii) $\hat{\vecom}_{\sst 0}$ is along the $z$-axis, and (iii) $\langle \vecf\rangle=0$ due to the periodicity of the particle's velocity,
one obtains 
\begin{equation}
\langle \vecv^{\sst (1)} \rangle = \langle v^{\sst (1)} \rangle \,\vecx = \frac{1}{9}
\Re\left[\hat{\omega}_{\sst{0}} \, \hat{v}_{{\sst{0}},y}^\dag \, \mathcal{I}(\tlam)\right]\,\vecx,
\label{eq:prop_x_comp_but_final}
\end{equation}
or equivalently, as a function of the tilt angle amplitude, 
\begin{equation}
\langle \vecv^{\sst (1)} \rangle = - \frac{1}{9}\,\veps\,\Im\left[\hat{\theta}_{\sst{0}} \, 
\hat{v}_{{\sst{0}},y}^\dag \, \mathcal{I}(\tlam)\right]\,\vecx,
\label{eq:prop_x_comp_final}
\end{equation}
where $\Im(z)$ refers to the imaginary part of $z$.

Note that the ratio $\langle v^{\sst(1)} \rangle/\veps$, which is a function of the four dimensionless parameters
$\alpha$, $\beta$, $I$ and $\lambda$, does not depend on $\veps$. As a result, the leading order dimensionless 
mean velocity of the particle $\oll{\vecv}$ is the product of $\varepsilon\,\Reyn$ and of a dimensionless 
function of the four other parameters. The quantity $\langle v^{\sst(1)} \rangle/\veps$
is plotted in Fig.~\ref{fig:H_vs_lambda} for $\beta = 0.2$ and different
combinations $(I,\alpha)$. 

\subsection{\color{black} Asymptotic behaviour and reversal of the propulsion speed}
As shown on Fig.~\ref{fig:H_vs_lambda}, for large $\alpha$ or small $I$, a reversal of the direction of propulsion
(illustrated by the diagrams inserted in each sub-figure) can be observed 
at a finite value $\lambda^\star$ of the reduced frequency $\lambda$. This reversal in swimming direction is not the result of the difference in behaviour of the streaming flows at low and high frequencies, and is instead entirely due to a change by a factor of $\pi$ in the relative phase between translation and rotation in the viscous (i.e. $\mbox{Re}=0$) response of the forced sphere.

\begin{figure}
\begin{center}
\includegraphics[width=.8\textwidth]{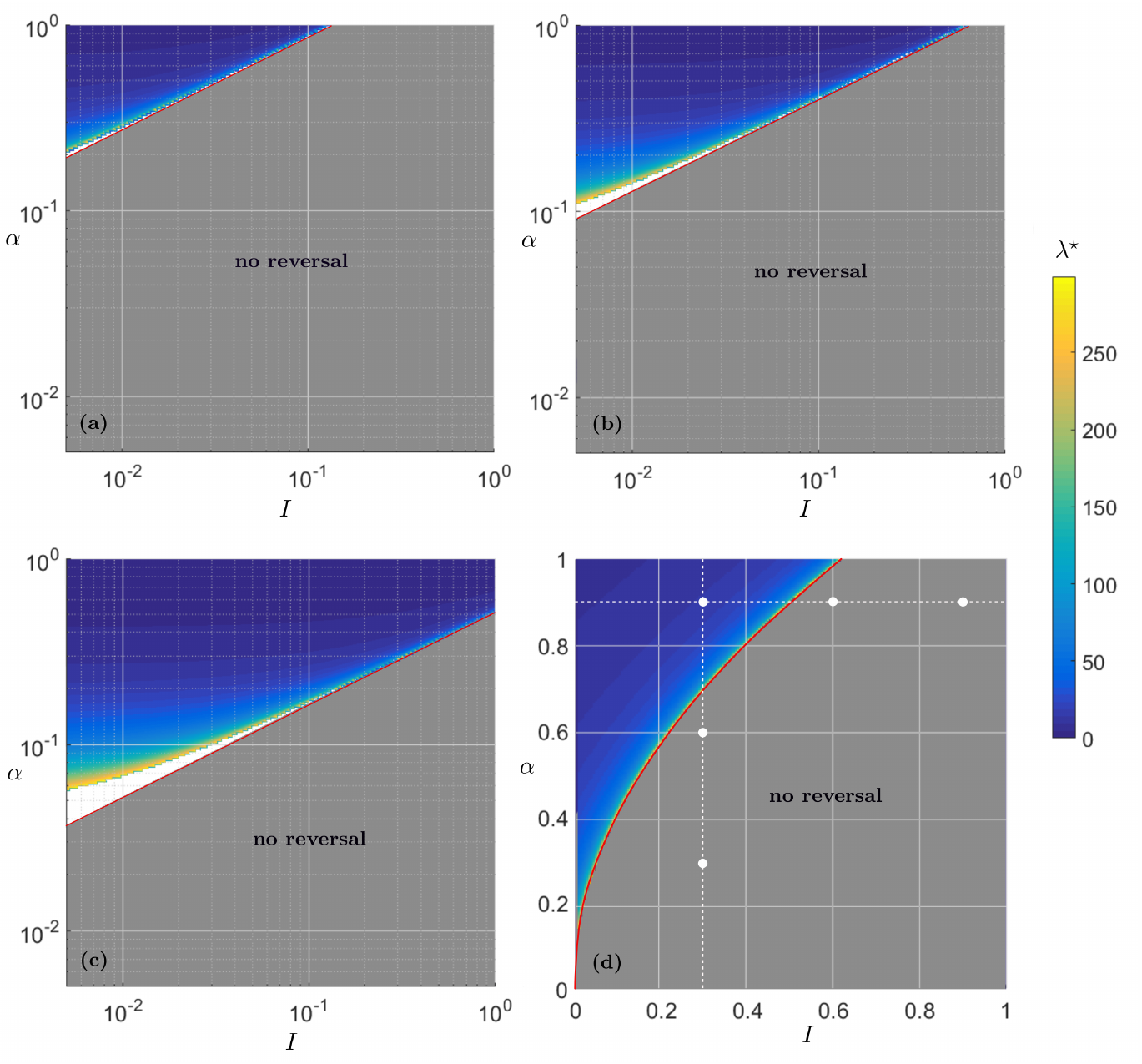}
\caption{Value $\lambda^\star$ corresponding to a reversal of the propulsion direction plotted in the plane $(I,\alpha)$
for three values of $\beta$. (a) $\beta = 0.05$ ; (b,d) $\beta = 0.2$ - figure (d) is the same as
(b) but plotted in a linear scale ; (c) $\beta = 0.6$.
For a fixed value of $\beta$, the frontier between the reversal and the non-reversal
regions is given by the equality case of Eq.~\eqref{eq:limit} (solid red line).
The cases considered in Fig.~\ref{fig:H_vs_lambda} are specified on (d) using white bullets.
The color bar (top right) holds for all the figures. The white zones in Figs (a), (b) and (c), where $\lambda^\star$ is
not computed, comes from the practical limitation in the numerical extraction of $\lambda^\star$ which tends to infinity in the vicinity of the transition (see text for further explanation).}
\label{fig:inversion}
\end{center}
\end{figure}
The variations of $\lambda^\star(\beta,I,\alpha)$ are plotted in Fig.~\ref{fig:inversion}, for
three different values of the density ratio $\beta$. In each case, the $(I,\alpha)$-plane is divided into two regions:
a first one where a reversal of the direction of propulsion can be observed at finite $\lambda$, and another one, where
the direction of propulsion does not depend on $\lambda$ (in the latter case, the sphere always propels
with the light end ahead). The limit between the two regions (i.e. a 
criterion for existence of the reversal in swimming direction between small and large $\lambda$) can be
obtained by deriving the asymptotic behaviour of $\langle v^{\sst(1)}\rangle$ at small and large $\lambda$. Substituting the result of Eqs.~\eqref{eq:asymp_small_lambda}, \eqref{eq:asymp_large_lambda} and \eqref{eq:asymp_I} into Eq.~\eqref{eq:prop_x_comp_final}, one obtains
\begin{gather}
\langle v^{\sst(1)}\rangle\sim\frac{\varepsilon\alpha\lambda^4(\beta-1)}{972\beta^2}\;\;\mbox{for}\;\;\lambda\rightarrow 0,\\
\langle v^{\sst(1)}\rangle\sim\frac{10\varepsilon \alpha\beta [2I(\beta-1)+5\alpha^2\beta]}{3[4I+\beta(2I+5\alpha^2)]^2}\;\;\mbox{for}\;\;\lambda\rightarrow \infty.
\end{gather}
A change in swimming direction between the $\lambda\ll 1$ and $\lambda \gg 1$ limits therefore requires $\beta-1$ and $2I(\beta-1)+5\alpha^2\beta$ to have opposite signs, or equivalently
\begin{equation}
0 \leq \frac{2(1-\beta)}{5\beta}\leq \frac{\alpha^2}{I}.\label{eq:limit}
\end{equation}
This is consistent with the results shown in Fig.~\ref{fig:inversion} where 
the red line corresponds to the equality case above.
The presence of a white zone in Figs.~\ref{fig:inversion}a-c, where $\lambda^\star$ is
not computed, comes from the practical limitation in the numerical
extraction of $\lambda^\star$ which tends to infinty in the vicinity of the transition region (red line).
This region would be reduced if the upper bound of the research interval in
$\lambda$ was enlarged. This has been verified for the value $\beta = 0.05$, for 
which the white zone barely exists. Note that the reversal is only possible 
if $\beta\leq 1$ (i.e. the particle must be heavier than the fluid on average) and 
if a sufficiently large inhomogeneity exists (as measured by $\alpha$). 
 
\section{Physical discussion and orders of magnitude\label{sec:OM}}

The dimensional form of Eq.~\eqref{eq:prop_x_comp_final} is
\begin{equation}
\oll{V} = \xi \varpi\,\oll{v} = -\frac{\xi^3\varpi^2}{9 \nu}\,\Im\left[\hat{\theta}_{\sst{0}} 
\, \hat{v}_{\sst{0}}^\dag \, \mathcal{I}(\tlam)\right],
\end{equation}
where it should be noted that the radius of the particle only appears through $\lambda$ (and not in the pre-factor).

Based on a mean value of the quality factor of the acoustic cavity $Q\sim 300$ \citep[][]{Bruus2012}
 and a typical displacement of the piezoelectric wall $\ell\sim 0.1\,$nm, a maximum value for the
displacement amplitude at the pressure node can be estimated as $\xi= 2Q\ell/\pi \sim 19\,$nm. 
For a typical particle radius $a = 0.5\;\mu$m and forcing frequency of $4\,$MHz, 
corresponding respectively to $\veps = 0.038$ and $\lambda \simeq 2.5$,
a value of $|\langle v^{\sst(1)} \rangle|/\veps \sim 0.01$ is a reasonable 
estimate of the particle's dimensionless velocity
(see Fig.~\ref{fig:H_vs_lambda}) and one obtains dimensionally
\begin{equation}
\oll{V} \sim 44\,\mbox{$\mu$m$\,$s$^{-1}$},
\end{equation}
which is consistent with the values reported by Ref.~\cite{Ahmed2016}. A quality factor of $10^3$ (upper
bound measured in standard acoustic resonators, see again\citep[][]{Bruus2012})
would have led to a propulsion velocity $\oll{V} \simeq 1.6\,$mm$\,$s$^{-1}$, which is much larger
than the values reported by Refs.~\citet{Ahmed2016} or \citet{Wang2012} (the latter reports a maximum value of 200
$\mu$m$\,$s$^{-1}$), but is not inconsistent with the velocities measured
by Ref.~\cite{Kaynak2017}. A quality factor of $10^2$ (lower bound measured in standard acoustic resonators)
would yield $\oll{V} \simeq 1.6\,\mu$m$\,$s$^{-1}$. 

In brief, even if the orders of magnitude of propulsion
speed produced by the model are not irrelevant to the measurements, performing a quantitative
comparison remains difficult because (i) the spherical geometry of our model noticeably departs
from the experimental geometry depicted in Refs.~\cite{Wang2012} and \cite{Ahmed2016} and (ii) the quality
factor of the experimental acoustic cavities used by Refs.~\cite{Wang2012} and \cite{Ahmed2016} is not known, 
whereas it critically impacts the estimate of the velocity.
\begin{figure}
\begin{center}
\includegraphics[width=.55\textwidth]{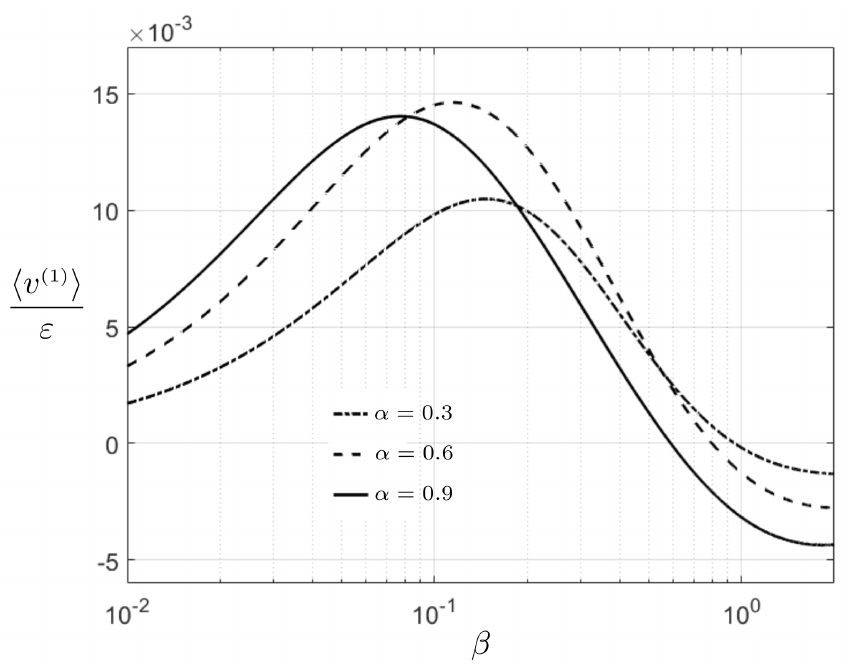}
\caption{(a) Ratio $\langle v^{\sst(1)} \rangle/\veps$ as a functions
of $\beta$ for $\lambda = 2.5$, $I = 0.9$ and different values of $\alpha$.}
\label{fig:beta}
\end{center}
\end{figure}
The profile of the quantity $\langle v^{\sst(1)} \rangle/\veps$ with respect to the density ratio $\beta$,
for $\lambda = 2.5$, $I = 0.9$ and different values of the parameter $\alpha$
is plotted in Fig.~\ref{fig:beta}. As mentioned by Ref.~\cite{Ahmed2016}, homogeneous 
Rhodium rods ($\beta = 0.081$) were faster than heavier golden ones ($\beta = 0.052$),
an observation which is again consistent with the values presented in the figure. 

A final practical yet fundamental remark must be made regarding the zero mean value of the tilt angle.
Indeed, we assumed here that the tilt angle varied periodically around the value $\theta = 0$ in the 
permanent regime (no angular drift).
This assumption is not \emph{a priori} fully-justified since the radiation
pressure on a sphere has no obvious orientation effect. In contrast, a near-sphere
or an ellipsoid will orient itself in such a way that, on average, its major axis would
lie in the zero-pressure plane of the wave. Therefore, the present calculation 
can be seen as the leading order calculation of the acoustic propulsion of a non-homogeneous 
near-sphere, since a slight alteration of the shape would not modify the propulsion speed obtained
at leading order for a non-homogeneous sphere.\\

\section{Conclusion}
\label{sec:conclusions}

We present here a full derivation of the acoustic propulsion speed of a non-homogeneous rigid
sphere. Unlike previous studies which generally rely on a numerically-integrated result, the final result
obtained by means of the inertial version of the Lorentz reciprocal theorem is integrated analytically. 
The problem is ruled by five independent dimensionless 
parameters: the in-homogeneity ratio or imbalance distance  $\alpha$, 
the fluid/solid density ratio $\beta$, the dimensionless
{\color{black} moment of inertia} $I$, the dimensionless forcing 
amplitude $\veps$ and the reduced frequency $\lambda^2$. 
For a given density ratio $\beta\leq 1$,  a limit value $\lambda^\star$ of the parameter $\lambda$ 
may exist such that propulsion takes place in different directions at low frequency ($\lambda<\lambda^\star$) 
and high frequency ($\lambda>\lambda^\star$). A necessary and sufficient condition for the existence
of a reversal in the propulsion direction for varying $\lambda$ was 
obtained as  $0 \leq (2/5)[(1-\beta)/\beta] \leq \alpha^2/I $.

The trends of the propulsion speed as a function of $\lambda$ as well as the possible existence of a 
change in propulsion direction for $\lambda=\lambda^\star$  are fully consistent with the results
published by Ref.~\cite{Collis2017}. As expected, in a case where the reversal value $\lambda^\star$ does exist
(see Fig.~\ref{fig:H_vs_lambda}), propulsion occurs at low frequency (small $\lambda$) in the direction 
of the lighter part of the sphere ({\color{black} center of mass} behind the centroid) whereas at higher frequency 
(large $\lambda$), the inhomogeneous sphere propels with the {\color{black} center of mass} ahead.
The dependence of the propulsion speed amplitude upon the density ratio is non monotonous and at high
mean densities (typically for $\beta < 0.1$), light particles propel faster than heavier ones.
Yet, one should be cautious in connecting this result
to the observation reported by Ref.~\cite{Ahmed2016} on density effects. Indeed, Ref.~\cite{Ahmed2016} report that
lighter rods propel faster than denser ones, but the rods also display a geometric asymmetry 
which could play a central role as well. 
In order to test more thoroughly the model, dedicated experiments
performed using low aspect ratio solid particles with controlled density inhomogeneities would be enlightening.

\section*{Acknowledgments}
This project has received funding from the
European Research Council (ERC) under the European Union's Horizon
2020 research and innovation programme under Grant Agreement 714027 (SM). 
The authors also acknowledge insightful discussions with K. Lippera, M. Benzaquen
and E. Lauga on the problem.

\appendix 
\section{Definition of the coefficients \boldmath$A$, $A^\star$, $B$, $B^\star$, $C$ and $C^\star$ \label{app:viscous_steady_harmonic}}
The full expressions of the coefficients $A$, $B$ and $C$ of the unsteady harmonic Stokes flow in Eq.~\eqref{eq:harmonic_viscous_field} are given by 
\begin{align}
A(r)&=\frac{3}{2\tlam^2r^3}\left[(1+\tlam r+\tlam^2r^2)\expo^{\tlam(1-r)}-1-\tlam-\frac{\tlam^2}{3}\right]
\label{eq:def_const_A}\\
B(r)&=\frac{3}{2\tlam^2r^3}\left[3+3\tlam+\tlam^2-(3+3\tlam r+\tlam^2r^2)\expo^{\tlam(1-r)}\right]
\label{eq:def_const_B}\\
C(r)&=\frac{\expo^{\tlam(1-r)}(1+\tlam r)}{(1+\tlam)r^2}
\label{eq:def_const_C}
\end{align}
where $\tilde\lambda=\lambda\mathrm{e}^{\mathrm{i}\pi/4}$, and the 
corresponding coefficients $A^\star$, $B^\star$ and $C^\star$ of the auxiliary steady Stokes flow in Eq.~\eqref{eq:steady_viscous_field} are given by
\begin{align}
A^\star(r)=\frac{3}{4r}+\frac{1}{4r^3},\qquad 
B^\star(r)=\frac{3}{4r}-\frac{3}{4r^3},\qquad
C^\star(r)=\frac{1}{r^2}\cdot
\label{eq:def_const_Cstar}
\end{align}

\section{Integration of the streaming term \boldmath$\mathcal{H}$ in the harmonic case \label{app:I}}

We note here $J(r)$ the integrand in the right-hand side of Eq.~\eqref{eq:I}, namely
\begin{equation}
J(r) = 
\left[A^\star\left(A^\dag C'+B^\dag C'+\frac{2A^\dag C}{r}\right)
+\frac{B^\star(A^\dag-B^\dag) C}{r}\right]r^2
\end{equation}
which can be rewritten explicitly as %
\begin{align}
J(r)=\frac{1}{4\tlam^{\dag 2}(1+\tlam)}\Bigg[&\expo^{\tlam(1-r)}(\tlam^{\dag 2} 
+ 3\tlam^\dag+3)\left(-\frac{3\tlam^2}{r^3}-\frac{15\tlam}{r^4}+\frac{3\tlam}{r^6}+\frac{3}{r^7}
-\frac{\tlam^2+15}{r^5}\right)\nonumber\\
&+3\expo^{(\tlam+\tlam^\dag)(1-r)}\Bigg(\frac{3\tlam\tlam^\dag(\tlam+2\tlam^\dag)}{r^2}
+\frac{3(\tlam^2+5\tlam\tlam^\dag+2\tlam^{\dag 2})}{r^3}\nonumber\\
&+\frac{\tlam^2\tlam^\dag-2\tlam\tlam^{\dag 2}+15\tlam+15\tlam^\dag}{r^4}\nonumber\\
&+\frac{\tlam^2-3\tlam\tlam^\dag-2\tlam^{\dag 2}+15}{r^5}-\frac{3(\tlam+\tlam^\dag)}{r^6}
-\frac{3}{r^7}\Bigg)\Bigg]
\end{align}
So that $\mathcal{I}(\tlam)=\int_1^\infty J(r)\mathrm{d}r$ is obtained analytically as 
\begin{align}
\mathcal{I}(\tlam)=\frac{1}{4\tlam^{\dag 2}(1+\tlam)}
\Big\{&(\tlam^{\dag 2}+3\tlam^\dag+3)\big[-3\tlam^2I_3-15\tlam I_4+3\tlam I_6+3
I_7-(\tlam^2+15)I_5\big]\nonumber\\[-2mm]
& + 3\big[3\tlam\tlam^\dag(\tlam+2\tlam^\dag)\tilde{I}_2+3(\tlam^2+5\tlam\tlam^\dag + 
2\tlam^{\dag 2})\tilde{I}_3\nonumber\\
& +(\tlam^2\tlam^\dag-2\tlam\tlam^{\dag 2}+15\tlam+15\tlam^\dag)\tilde{I}_4\nonumber\\
& +(\tlam^2-3\tlam\tlam^\dag-2\tlam^{\dag 2}+15)\tilde{I}_5-3(\tlam+\tlam^\dag)
\tilde{I}_6-3\tilde{I}_7\big]\Big\}
\end{align}
with
\begin{align}
I_1 & =\int_1^\infty\frac{\expo^{\tlam(1-r)}}{r}\,dr=\expo^{\tlam}E_1(\tlam)\\
\tilde{I}_1 & =\int_1^\infty\frac{\expo^{(\tlam+\tlam^\dag)(1-r)}}{r}\,dr
= \expo^{(\tlam+\tlam^\dag)}E_1(\tlam+\tlam^\dag)\\
I_n & =\int_1^\infty\frac{\expo^{\tlam(1-r)}}{r^n}\,dr = \frac{1}{n-1}(1-\tlam I_{n-1})\\
\tilde{I}_n & =\int_1^\infty\frac{\expo^{(\tlam+\tlam^\dag)(1-r)}}{r^n}\,dr
= \frac{1}{n-1}[1-(\tlam+\tlam^\dag) \tilde{I}_{n-1}]
\end{align}
which are well defined since $\tlam$ has positive real part. In the above equation, $E_1(z)$ 
is the exponential integral \citep{abramowitz1964}. Using these results, one obtains $\mathcal{I}$ analytically for any $\lambda$:
\begin{align}
\mathcal{I}(\tlam)=&\frac{1}{64\tlam^\dag(1+\tlam)}\big[(\tlam^\dag+3)\tlam^5 
-\tlam^\dag \tlam^4 + (2\tlam^\dag - 18)\tlam^3 \nonumber\\
& -6\tlam^{\dag 2}\tlam^2 +(-3\tlam^{\dag 3} + 6\tlam^{\dag 2} - 16\tlam^\dag)\tlam 
+ 3\tlam^{\dag 4} - 3\tlam^{\dag 3} - 48\tlam^{\dag 2} - 16\tlam^\dag\big]\nonumber\\
& -\frac{\expo^{\tlam} E_1(\tlam)}{64\tlam^{\dag 2}(1+\tlam)}
\big[(\tlam^{\dag 2}+3\tlam^\dag +3)(\tlam^6 -6\tlam^4)\big]\nonumber\\
& +\frac{3\expo^{\tlam+\tlam^\dag} E_1(\tlam+\tlam^\dag)}{64\tlam^{\dag 2}(1+\tlam)}
(\tlam+\tlam^\dag)(\tlam-\tlam^\dag)[\tlam^4-2\tlam^2\tlam^{\dag 2}
-6\tlam^2+\tlam^{\dag 4}-18\tlam^{\dag 2}].
\end{align}
Real, imaginary part and phase of $\mathcal{I}$ are presented in Fig.~\ref{fig:I}. The asymptotic forms of $\mathcal{I}$ at small and large $\lambda$, Eq.~\eqref{eq:asymp_I}, are obtained using the following asymptotic limits~\citep{abramowitz1964}
\begin{equation}
E_1(z\rightarrow 0)\sim -\ln z-\gamma ,\qquad E_1(z\rightarrow\infty)\sim\frac{\mathrm{e}^{-z}}{z}\cdot
\end{equation}


\end{document}